\documentclass[12pt,a4paper]{article}

\usepackage{amsmath, amssymb, amsfonts}
\usepackage{bm}
\usepackage{physics}
\usepackage{geometry} 
\usepackage{graphicx}
\usepackage{hyperref}
\begin{document}

\title{On the Connection Between Chaos-Assisted Tunneling and Coherent Destruction of Tunneling}
\author{\bf Sumita Datta $^{1,2}$\\
$^1$ Department of Pure and Applied Mathematics, Alliance University,\\ Bengaluru 562 106, India\\
$^2$ Department of Physics, University of Texas at Arlington,\\Texas 76019, USA\\}
\maketitle
\begin{abstract}
The interplay between classical chaos and quantum tunneling is examined in driven nonlinear systems, with emphasis on how semiclassical phase–space structures influence purely quantum transport phenomena. We show that, in the presence of external driving and stochastic perturbations, tunneling rates acquire an activated form determined by effective classical barriers, providing a transparent link between chaotic dynamics and quantum tunneling. Within this framework, chaos-assisted tunneling and coherent destruction of tunneling emerge as closely related manifestations of the same underlying phase–space restructuring and interference effects induced by driving. The results offer a unified perspective on tunneling control in nonintegrable systems and remain relevant for modern studies of driven quantum dynamics and decoherence-resistant transport.
\end{abstract}
\newpage
\section{Introduction}

The problem of relating classical chaotic dynamics to quantum evolution has long occupied a central position in mathematical physics, particularly in the context of semiclassical analysis and the correspondence principle \cite{Gutzwiller,Ozorio,Keller}. Classical chaos is associated with the breakdown of integrability, exponential instability of trajectories, and complex invariant structures in phase space, whereas quantum dynamics is governed by linear unitary evolution on Hilbert space. Understanding how classical phase--space features are encoded in quantum spectra, eigenstates, and transition processes remains a fundamental challenge, especially in nonintegrable and driven systems \cite{Haake,Stoeckmann}.

One of the most delicate manifestations of this quantum--classical interplay arises in tunneling phenomena. Quantum tunneling has no direct classical analogue, yet its rate and qualitative behavior can depend sensitively on the underlying classical dynamics. Early semiclassical studies demonstrated that tunneling amplitudes may be strongly influenced by nonlinear resonances and invariant manifolds, leading to the concept of \emph{dynamical tunneling} \cite{DavisHeller,LinBallentine}. In systems with mixed regular--chaotic phase space, tunneling between classically disconnected regular regions can be mediated by chaotic states, resulting in large fluctuations of tunneling rates and strong sensitivity to system parameters. This mechanism, commonly referred to as chaos--assisted tunneling, has been explored both numerically and analytically in a variety of nonintegrable models \cite{TomsovicUllmo,BohigasTomsovic}.

A complementary line of investigation concerns the influence of time--periodic driving on tunneling dynamics. From a Floquet--theoretic perspective, external modulation leads to a restructuring of the quasi--energy spectrum, and tunneling suppression may occur when destructive interference conditions are satisfied. This effect, known as coherent destruction of tunneling, was originally identified in driven double--well systems and interpreted in terms of quasi--energy degeneracies and symmetry properties of the effective Floquet Hamiltonian \cite{GrossmannDittrich,GrifoniHanggi}. Subsequent work has clarified the robustness of this phenomenon and its dependence on driving amplitude and frequency \cite{HolthausReview}. While often presented as a purely quantum interference effect, coherent destruction of tunneling also admits a semiclassical interpretation in terms of the modification of classical phase--space transport under periodic driving.

The role of stochastic perturbations further enriches the picture. In classical dynamics, noise leads to activated escape over dynamical barriers, with rates determined by large--deviation principles and effective action functionals rather than static potential heights \cite{Kramers,FreidlinWentzell}. Quantum mechanically, noise introduces decoherence and energy diffusion, altering spectral properties and transition probabilities \cite{CaldeiraLeggett}. In the semiclassical limit, it has been shown that quantum transition rates in noisy or driven systems can acquire Arrhenius--type forms, with exponents governed by classical actions \cite{DykmanKrivoglaz,DykmanSmelyanskiy}. These results suggest that quantum tunneling and classical activation may be viewed as different manifestations of a common underlying large--deviation structure.

In the present work, we develop this perspective in the context of driven nonlinear systems exhibiting mixed regular and chaotic dynamics. By formulating quantum transition rates in a semiclassical framework and explicitly incorporating the effects of periodic driving and stochastic perturbations, we show how tunneling rates reduce, in appropriate limits, to activated expressions controlled by effective classical barriers. Within this formulation, chaos--assisted tunneling and coherent destruction of tunneling emerge as closely related consequences of phase--space restructuring and spectral reorganization induced by driving. Our analysis provides a unified mathematical framework connecting semiclassical chaos, quantum tunneling, and noise--induced transport, and clarifies the mechanisms by which classical dynamical features influence quantum transition processes. In the present context another work worthmentioning is \cite{KuvshinovKuzminPiatrou}. 
\section{Origin of Chaos-Assisted Tunneling}

The deterministic Hamiltonian considered in this work,
\begin{equation}
H^{0}(t)=\frac{p^{2}}{2m}+Rx^{4}-Qx^{2}-\lambda x\cos(\omega t),
\end{equation}
contains two distinct ingredients: a nonlinear double--well potential and an external time--periodic driving term. The static double--well Hamiltonian,
\[
H_{\mathrm{dw}}=\frac{p^{2}}{2m}+Rx^{4}-Qx^{2},
\]
is nonlinear but integrable, possessing a single degree of freedom and a regular phase--space structure. In the absence of driving, the classical dynamics is nonchaotic and tunneling occurs directly between regular regions of phase space.

The origin of chaos in the present system can be traced to the time--periodic driving term $-\lambda x\cos(\omega t)$. When combined with the nonlinear potential, this term renders the classical dynamics nonintegrable by effectively increasing the dimensionality of phase space. As a result, nonlinear resonances and resonance overlap occur, leading to the destruction of invariant tori and the formation of chaotic regions, particularly near separatrices. Thus, chaos arises from the interplay between nonlinearity and driving, rather than from either ingredient alone.

Quantum mechanically, the presence of classically chaotic regions has a profound impact on tunneling dynamics. Eigenstates associated with chaotic motion are typically delocalized over large regions of phase space and can overlap significantly with states localized in both wells. This provides indirect pathways for tunneling between symmetry--related regular states, mediated by chaotic intermediate states. In this sense, tunneling is not a direct process but is assisted by chaotic dynamics, giving rise to what is commonly referred to as chaos--assisted tunneling.

Within the harmonic--oscillator representation employed here, this mechanism manifests itself through the combined action of the nonlinear static terms and the time--periodic driving, which together produce strong mixing of basis states and a dense set of intermediate quasi--energy levels. The resulting enhancement or suppression of tunneling rates is therefore controlled by the structure of the underlying classical phase space. In regimes where driving reorganizes phase space so as to suppress chaotic transport, the same mechanism leads to coherent destruction of tunneling, highlighting the close connection between chaos--assisted tunneling and coherent control of quantum transport.
\newpage
\section{Noise-induced Transition Rate in the Driven Double-Well Oscillator}

We begin with the total potential
\begin{equation}
V(x,t) = Rx^{4} - Q x^{2} - \lambda x \cos(\omega t) + f(t),
\end{equation}
where where $Q, R>0$ define a symmetric double-well potential, and the time-dependent driving term $-\lambda x \cos(\omega t)$ provides a weak periodic perturbation and
$f(t)$ is a Gaussian white noise satisfying
\begin{equation}
\langle f(t) \rangle = 0, \qquad 
\langle f(t) f(t') \rangle = D\,\delta(t-t').
\end{equation}
Following Lin and Ballentine [Phys. Rev. A \textbf{45}, 3637 (1992)], the Hamiltonian matrix elements (their Eq.~(2.5)) are written in the harmonic oscillator basis $\{|n\rangle\}$ as
\begin{equation}
\begin{aligned}
H^{0}_{mn}(t)
&=
\hbar\omega_{0}\!\left(n+\tfrac12\right)\delta_{mn}
\\[1ex]
&\quad
-\frac{Q\hbar}{2m\omega_{0}}
\Big[
\sqrt{(n+1)(n+2)}\,\delta_{m,n+2}
+(2n+1)\,\delta_{mn}
+\sqrt{n(n-1)}\,\delta_{m,n-2}
\Big]
\\[1ex]
&\quad
+\frac{R\hbar^{2}}{4m^{2}\omega_{0}^{2}}
\Big[
\sqrt{(n+1)(n+2)(n+3)(n+4)}\,\delta_{m,n+4}
\\
&\qquad\qquad
+2(2n+3)\sqrt{(n+1)(n+2)}\,\delta_{m,n+2}
\\
&\qquad\qquad
+3(2n^{2}+2n+1)\,\delta_{mn}
\\
&\qquad\qquad
+2(2n-1)\sqrt{n(n-1)}\,\delta_{m,n-2}
\\
&\qquad\qquad
+\sqrt{n(n-1)(n-2)(n-3)}\,\delta_{m,n-4}
\Big]
\\[1ex]
&\quad
-\lambda\cos(\omega t)\,
\sqrt{\frac{\hbar}{2m\omega_{0}}}
\Big[
\sqrt{n+1}\,\delta_{m,n+1}
+\sqrt{n}\,\delta_{m,n-1}
\Big].
\end{aligned}
\tag{2.5}
\end{equation}

We now add the stochastic contribution due to $f(t)$, which being spatially uniform contributes only to the diagonal terms:
\begin{equation}
H_{mn}(t) = H_{mn}^{(0)}(t) + f(t)\,\delta_{mn}.
\end{equation}
The deterministic part $H_{mn}^{(0)}(t)$ is identical to that in Lin and Ballentine, while the random part adds a fluctuating phase to all basis states.

\subsection{Dephasing rate produced by the diagonal white noise}

In the interaction picture with respect to $H_{0}$, the noise Hamiltonian is
\begin{equation}
H_{\text{noise}}(t) = f(t)\,\mathbb{I}.
\end{equation}
The random phase accumulated by an eigenstate over time $t$ is
\begin{equation}
\phi(t) = \frac{1}{\hbar}\int_{0}^{t} f(s)\,ds.
\end{equation}
Because $\langle f(s) f(s') \rangle = D \delta(s-s')$, the variance of this random phase is
\begin{equation}
\mathrm{Var}[\phi(t)] = 
\frac{1}{\hbar^{2}} \int_{0}^{t} \! ds \int_{0}^{t} \! ds' \,
\langle f(s)f(s') \rangle
= \frac{D}{\hbar^{2}} t .
\end{equation}
Averaging over realizations of $f(t)$ gives the decay of coherence between energy eigenstates:
\begin{equation}
\Big\langle e^{-i(\phi(t)-\phi(0))} \Big\rangle
= \exp\!\left[-\frac{1}{2}\,\mathrm{Var}[\phi(t)]\right]
= \exp\!\left[-\frac{D}{2\hbar^{2}}\,t\right].
\end{equation}
Thus the uniform white noise produces an exponential decay of off-diagonal density-matrix elements with a \emph{decoherence rate}
\begin{equation}
\boxed{\gamma = \frac{D}{2\hbar^{2}}.}
\end{equation}

\subsection{Transition amplitude due to the periodic driving term and white noise}

We expand the quantum state in the eigenbasis $\{|k\rangle\}$ of the unperturbed Hamiltonian $H_0$,
\begin{equation}
|\psi(t)\rangle=\sum_k c_k(t)\,e^{-iE_k t/\hbar}\,|k\rangle ,
\end{equation}
where $E_k$ are the eigenvalues of $H_0$. Substitution into the time--dependent Schr\"odinger equation yields the exact interaction--picture evolution equation
\begin{equation}
i\hbar\,\dot c_n(t)
=
\sum_k
\langle n|V(t)|k\rangle
\,e^{i\omega_{nk}t}\,
c_k(t),
\qquad
\omega_{nk}=\frac{E_n-E_k}{\hbar},
\label{eq:exact_ck}
\end{equation}
with the time--dependent perturbation
\begin{equation}
V(t)=-\lambda\,x\cos(\omega t).
\end{equation}

\paragraph{First--order perturbative approximation.}
We consider transitions from an initial state $|m\rangle$, so that the initial conditions are
\begin{equation}
c_k(0)=\delta_{km}.
\end{equation}
In first--order time--dependent perturbation theory, the coefficients $c_k(t)$ on the right--hand side of Eq.~\eqref{eq:exact_ck} are replaced by their zeroth--order values,
\begin{equation}
c_k(t)=\delta_{km}+\mathcal{O}(\lambda).
\end{equation}
This replacement is consistent because corrections to $c_k(t)$ are already of order $\lambda$, and retaining them would generate contributions of order $\lambda^2$ to the amplitude.

With this approximation, Eq.~\eqref{eq:exact_ck} reduces to
\begin{equation}
i\hbar\,\dot c_n^{(1)}(t)
=
\langle n|V(t)|m\rangle
\,e^{i\omega_{nm}t},
\label{eq:first_order_eq}
\end{equation}
where the superscript $(1)$ denotes first order in $\lambda$.

\paragraph{Integration and transition amplitude.}
Integrating Eq.~\eqref{eq:first_order_eq} from $0$ to $t$ gives the first--order transition amplitude
\begin{equation}
c_n^{(1)}(t)
=
-\frac{i}{\hbar}
\int_{0}^{t} dt'\,
\langle n|V(t')|m\rangle
\,e^{i\omega_{nm}t'} .
\label{eq:transition_amplitude_general}
\end{equation}
Substituting the explicit form of the perturbation,
\begin{equation}
c_n^{(1)}(t)
=
\frac{i\lambda}{\hbar}
\langle n|x|m\rangle
\int_{0}^{t} dt'\,
\cos(\omega t')\,e^{i\omega_{nm}t'} .
\label{eq:transition_amplitude_cos}
\end{equation}

\paragraph{Consistency of the approximation.}
The explicit coefficient $c_k(t)$ does not appear in
Eqs.~\eqref{eq:first_order_eq}--\eqref{eq:transition_amplitude_cos}
because it has been replaced by its initial value $\delta_{km}$ as part of the first--order approximation. Retaining the full time dependence of $c_k(t)$ would generate higher--order corrections in $\lambda$ and is therefore inconsistent at this order. The transition probability, obtained from $|c_n^{(1)}(t)|^2$, is already of order $\lambda^2$.

Now on we will drop the superscript of $c_k(t)$ as it is implied that our matrixelements are calculated using the first order perturbation theory only.  
Thus, the disappearance of $c_k(t)$ in the expression for the transition amplitude is a direct consequence of first--order time--dependent perturbation theory, in which
\begin{equation}
c_k(t)\;\longrightarrow\; c_k(0)=\delta_{km}
\end{equation}
inside the time integral. This approximation is mathematically controlled and forms the standard starting point for the derivation of transition rates and Fermi's golden rule.

Let the system be initially in the eigenstate $\ket{n}$.
The first-order time-dependent perturbation theory gives the transition amplitude to $\ket{m}$ as
\begin{equation}
c_{n\leftarrow m}(t)
= -\frac{i}{\hbar} \int_0^t \! dt'\, 
\langle m | V(t') | n \rangle
\,e^{i\omega_{mn}t'},
\label{eq:amp}
\end{equation}
where $\omega_{mn} = (E_m^{(0)} - E_n^{(0)})/\hbar$ and
\begin{equation}
V(t) = -\lambda x \cos(\omega t) + f(t).
\end{equation}
\begin{align}
c_{n\leftarrow m}(t)
&= \frac{i\lambda}{\hbar} \langle m|x|n\rangle
  \int_0^t \! dt'\, \cos(\omega t') e^{i\omega_{mn}t'}
 - \frac{i}{\hbar} \int_0^t \! dt'\, f(t') e^{i\omega_{mn}t'} \delta_{mn}.
\end{align}

The first term describes transitions due to the coherent drive $-\lambda x \cos(\omega t)$, and the second term describes random phase modulation due to the stochastic force.
The transition probability from $\ket{n}$ to $\ket{m}$ is
\begin{equation}
W_{mn}(t) =  |c_{n\leftarrow m}(t)|^{2} 
\end{equation}

Since the noise term contributes only for $m=n$ (diagonal), it does not directly induce transitions.
For $m\neq n$, the transition probability is governed by the driven term:
\begin{align}
W_{mn}(t)
&= \frac{\lambda^2}{\hbar^2} |\langle m|x|n\rangle|^2
   \left| \int_0^t \! dt'\, \cos(\omega t') e^{i\omega_{mn}t'} \right|^2.
\end{align}

The time-dependent perturbation due to peridic driving term is
\begin{equation}
H'(t) = -\lambda \hat{x} \cos(\omega t)
       = -\frac{\lambda}{2}\,\hat{x}\,
         \big(e^{i\omega t}+e^{-i\omega t}\big).
\end{equation}
In the interaction picture, the transition amplitude from $\lvert m\rangle$ to $\lvert n\rangle$ to first order in $\lambda$ because of just periodic driving term is
\begin{equation}
c_{n\leftarrow m}(t)
= -\frac{i}{\hbar}\frac{\lambda}{2}\,x_{nm}
  \int_{0}^{t} dt'\,
  \big(e^{i(\omega_{nm}+\omega)t'}+e^{i(\omega_{nm}-\omega)t'}\big)
  e^{-\gamma t'},
\end{equation}
where 
$x_{nm}=\langle n|\hat{x}|m\rangle$ are the position matrix elements
in the harmonic-oscillator basis.

For long times $t\gg 1/\gamma$, the time integral yields Lorentzian denominators:
\begin{equation}
\left|\int_{0}^{t} e^{i(\omega_{nm}-\omega)t'} e^{-\gamma t'} dt' \right|^{2}
\approx
\frac{1}{(\omega_{nm}-\omega)^{2}+\gamma^{2}}\;(\pi t)\gamma.
\end{equation}

Hence the transition \emph{rate}
$W_{m\to n}=\lim_{t\to\infty}|c_{n\leftarrow m}(t)|^{2}/t$ which is precisely the Fermi Golden rule becomes
\begin{equation}
\boxed{
W_{m\to n}
= \frac{\lambda^{2}|x_{nm}|^{2}}{2\hbar^{2}}\,
  \frac{\gamma}{(\omega_{nm}-\omega)^{2}+\gamma^{2}}
 +\frac{\lambda^{2}|x_{nm}|^{2}}{2\hbar^{2}}\,
  \frac{\gamma}{(\omega_{nm}+\omega)^{2}+\gamma^{2}}.
}
\end{equation}
On resonance $(\omega\approx\omega_{nm})$, the first term dominates.

It is sometimes convenient to write this using a broadened delta function
\begin{equation}
\delta_{\gamma}(\Delta)
=\frac{1}{\pi}\frac{\gamma}{\Delta^{2}+\gamma^{2}},
\end{equation}
so that
\begin{equation}
W_{m\to n}
=\frac{\pi\lambda^{2}}{2\hbar^{2}}|x_{nm}|^{2}\,
\big[\delta_{\gamma}(\omega_{nm}-\omega)
     +\delta_{\gamma}(\omega_{nm}+\omega)\big].
\end{equation}
\subsection{Comparison / correspondence with Kramers (classical) rate}

Kramers' classical escape rate for activated barrier crossing depends on the friction (dissipation), the temperature $T$ (noise strength), and the potential barrier height $\Delta V$. The \emph{qualitative} points of contact are these:

\subsubsection{Noise produces broadening / smearing of energy resonances.}
In the quantum calculation above the diagonal white noise produced a dephasing rate $\gamma=D/(2\hbar^{2})$. That broadening replaces sharp energy conservation $\delta(\omega_{nm}-\omega)$ by a Lorentzian of width $\gamma$. In the classical Kramers picture, thermal noise and friction determine the rate at which trajectories get enough energy to cross the barrier; the friction and noise set the timescale (prefactor) while $\exp(-\Delta V/k_{B}T)$ gives the exponential suppression. Thus in both pictures noise controls the \emph{rate prefactor} and the effective spectral overlap needed for escape.

\subsubsection{Semiclassical / many-level limit $\hbar\to0$.}
When levels become dense near the barrier energy, summing quantum transition rates over many final states and using the semiclassical expressions for matrix elements $|x_{nm}|^{2}$ and the density of states converts the quantum golden-rule expression into a classical transition (diffusion) rate across the separatrix. Formally,
\begin{equation}
W_{\text{escape,\,quantum}}
\sim
\int dE'\,\rho(E')\,\frac{\lambda^{2}}{\hbar^{2}}\,|x(E,E')|^{2}\,
\delta_{\gamma}(E'-E\pm\hbar\omega),
\end{equation}
and in the semiclassical limit $\rho(E')$ large and $|x(E,E')|^{2}$ related to classical Fourier components of $x(t)$ along trajectories, this integral becomes the classical rate of energy diffusion across the barrier. The noise-induced width $\gamma$ plays the role of the classical noise constant that sets diffusion across energies. (A standard semiclassical derivation is to express matrix elements by Fourier transforms of the classical coordinate along a trajectory and carry out the sum over final states --- that converts quantum spectral factors to classical power spectra.)

\subsubsection{Role of noise intensity $D$.}
In the classical Kramers picture the escape prefactor scales with the noise intensity (or temperature) in a way determined by whether the dynamics are underdamped or overdamped; in the quantum golden-rule picture increasing $D$ increases $\gamma$, which increases the Lorentzian overlap and thus can \emph{increase} the transition rate off resonance (while on exact resonance extreme broadening eventually reduces peak height). So both frameworks show that the noise intensity controls the effective escape prefactor and, in the appropriate limit and after a careful semiclassical averaging, the quantum formula \emph{matches} the classical Kramers escape law (same parametric dependence on noise level and on barrier properties) --- the detailed matching requires the usual semiclassical replacement of matrix elements by classical power spectra and accounting for the thermal noise spectrum (or its quantum analogue).

So the \emph{correspondence} is: the white noise enters quantum theory as a dephasing/broadening parameter $\gamma\propto D/\hbar^{2}$; in the semiclassical limit the quantum golden-rule sums map onto classical diffusion in energy which is the mechanism behind Kramers' escape.

So summarizing the classical Kramers escape rate describes the noise-induced transition of a particle over a potential barrier. In the present quantum model:

\begin{enumerate}
\item
The white-noise term introduces an energy-level broadening 
$\gamma=D/(2\hbar^{2})$, replacing the delta-function resonance condition 
by a Lorentzian profile.
\item
In the semiclassical limit where the level spacing near the barrier is small, the sum over quantum transition rates can be converted into an integral over energy:
\[
W_{\text{escape, quantum}}
\simeq \int dE'\,\rho(E')\,
\frac{\lambda^{2}}{\hbar^{2}}
|x(E,E')|^{2}\,
\delta_{\gamma}(E'-E\pm\hbar\omega).
\]
Using semiclassical relations between matrix elements and classical
Fourier components of $x(t)$ along trajectories, this expression reduces
to an energy-diffusion equation whose stationary solution reproduces
the classical Kramers rate, with $\gamma$ (or equivalently $D$)
playing the role of the classical noise intensity.
\item
Thus the noise-induced level broadening in quantum dynamics corresponds to the diffusion constant governing thermal activation in the classical limit.
\end{enumerate}


The transition amplitude in first-order perturbation theory is linear in $\lambda$, while the transition \emph{probability} or rate is quadratic:
\begin{equation}
c_{n\leftarrow m}^{(1)} \propto \lambda,
\qquad
P_{m\to n}=|c_{n\leftarrow m}^{(1)}|^{2}\propto\lambda^{2}.
\end{equation}
Therefore, to lowest nontrivial order, the quantum transition probability
is \emph{quadratic} in $\lambda$, not linear.


\begin{itemize}
\item The diagonal white-noise term adds pure dephasing with rate
\[
\boxed{\gamma = \dfrac{D}{2\hbar^{2}}.}
\]
\item The transition rate between states $|m\rangle$ and $|n\rangle$ due to the periodic driving is
\[
\boxed{
W_{m\to n}
= \frac{\lambda^{2}|x_{nm}|^{2}}{2\hbar^{2}}
  \left[
  \frac{\gamma}{(\omega_{nm}-\omega)^{2}+\gamma^{2}}
 +\frac{\gamma}{(\omega_{nm}+\omega)^{2}+\gamma^{2}}
  \right].
}
\]
\item On resonance ($\omega\simeq\omega_{nm}$), the transition rate is maximized and exhibits Lorentzian broadening governed by $\gamma$.
\end{itemize}	
\subsection{Derivation: reduction of \(W_{mn}^{(\mathrm{eff})}\) to the Dykman et al.\ form (Eq.~(6))}

We start from the effective quantum transition (escape) rate between states \(m\) and \(n\) obtained by averaging over the diagonal white noise (see main text):
\begin{equation}
W_{mn}^{(\mathrm{eff})}
\;=\;
\frac{\pi\lambda^{2}}{2\hbar^{2}}\,|x_{nm}|^{2}\,
\delta_{\gamma}(\omega_{nm}-\omega),
\qquad
\delta_{\gamma}(\Delta)\equiv\frac{1}{\pi}\frac{\gamma}{\Delta^{2}+\gamma^{2}},
\label{eq:Wmn_eff_repeat}
\end{equation}
with the noise-induced broadening
\begin{equation}
\gamma \;=\; \frac{D}{2\hbar^{2}}.
\label{eq:gammaD}
\end{equation}

The physical escape rate \(W_{\mathrm{esc}}\) from a metastable well is obtained by summing (or integrating, in the dense-spectrum limit) over final states near the barrier energy. In the semiclassical limit this sum may be replaced by an integral over energy:
\begin{equation}
W_{\mathrm{esc}} \simeq \int dE'\,\rho(E')\,
\frac{\pi\lambda^{2}}{2\hbar^{2}}\,|x(E,E')|^{2}\,
\delta_{\gamma}\!\big(\tfrac{E'-E}{\hbar}-\omega\big),
\label{eq:Wesc_integral}
\end{equation}
where \(\rho(E')\) is the density of states and \(|x(E,E')|^2\) denotes the semiclassical envelope of the matrix elements between states with energies \(E,E'\). Here \(E\) denotes the (initial) intrawell energy from which activation is considered.
\subsection{Derivation of the activated form from Eq.~(34)}

We start from the effective escape rate obtained after averaging over
the diagonal white noise,
\begin{equation}
W_{\mathrm{esc}}(E)
\simeq \int dE'\,\rho(E')\,
\frac{\pi\lambda^{2}}{2\hbar^{2}}\,|x(E,E')|^{2}\,
\delta_{\gamma}\!\Big(\frac{E'-E}{\hbar}-\omega\Big),
\label{eq:Wesc24_repeat}
\end{equation}
where
\[
\delta_{\gamma}(\Delta)=\frac{1}{\pi}\frac{\gamma}{\Delta^{2}+\gamma^{2}},
\qquad
\gamma=\frac{D}{2\hbar^{2}} .
\]

\paragraph{Step 1: Replace the broadened delta function.}
Using the explicit Lorentzian form of $\delta_\gamma$, Eq.~\eqref{eq:Wesc24_repeat}
becomes
\begin{equation}
W_{\mathrm{esc}}(E)
=
\int dE'\,\rho(E')\,
\frac{\lambda^{2}}{2\hbar^{2}}\,
|x(E,E')|^{2}
\frac{\gamma}{
\left(\tfrac{E'-E}{\hbar}-\omega\right)^{2}+\gamma^{2}}.
\label{eq:Wesc_lorentz}
\end{equation}

\paragraph{Step 2: Semiclassical form of matrix elements.}
In the semiclassical regime ($\hbar\to0$, dense spectrum),
the coordinate matrix elements admit the standard WKB/Fourier representation
\begin{equation}
|x(E,E')|^{2}
\simeq |X(E,\omega')|^{2},
\qquad
\omega'=\frac{E'-E}{\hbar},
\end{equation}
where $X(E,\omega')$ is the classical Fourier component of the coordinate
along a trajectory of energy $E$.
Thus Eq.~\eqref{eq:Wesc_lorentz} may be written as
\begin{equation}
W_{\mathrm{esc}}(E)
=
\frac{\lambda^{2}}{2\hbar}
\int d\omega'\,
\rho(E+\hbar\omega')\,
|X(E,\omega')|^{2}
\frac{\gamma}{(\omega'-\omega)^{2}+\gamma^{2}}.
\label{eq:Wesc_freq}
\end{equation}

\paragraph{Step 3: Dense-spectrum and smoothness approximation.}
Near the barrier, both $\rho(E')$ and $|X(E,\omega')|^{2}$ vary slowly
on the scale set by $\gamma$.
Hence they may be evaluated at the resonant value $\omega'=\omega$:
\begin{equation}
\rho(E+\hbar\omega')\,|X(E,\omega')|^{2}
\simeq
\rho(E+\hbar\omega)\,|X(E,\omega)|^{2}.
\end{equation}
The remaining integral is elementary:
\[
\int d\omega'\,
\frac{\gamma}{(\omega'-\omega)^{2}+\gamma^{2}}
= \pi .
\]
Therefore,
\begin{equation}
W_{\mathrm{esc}}(E)
\simeq
\frac{\pi\lambda^{2}}{2\hbar}\,
\rho(E+\hbar\omega)\,|X(E,\omega)|^{2}.
\label{eq:Wesc_energy}
\end{equation}

\paragraph{Step 4: Energy diffusion and escape.}
Escape over the barrier requires reaching energies
$E \gtrsim E_b$, where $E_b$ is the barrier top.
Noise induces slow diffusion in energy space.
The stationary probability density for intrawell energies satisfies
the Fokker--Planck equation
\begin{equation}
\frac{\partial P(E,t)}{\partial t}
=
\frac{\partial}{\partial E}
\left[
D\left(
\frac{\partial P}{\partial E}
+
\frac{1}{D}\frac{dU}{dE}P
\right)
\right],
\end{equation}
whose stationary solution is
\begin{equation}
P_{\mathrm{st}}(E)
\propto \exp\!\Big(-\frac{U(E)}{D}\Big).
\end{equation}

\paragraph{Step 5: Steepest-descent evaluation.}
The escape rate is obtained by weighting
Eq.~\eqref{eq:Wesc_energy} with $P_{\mathrm{st}}(E)$
and integrating over $E$:
\begin{equation}
W_{\mathrm{esc}}
=
\int dE\,
P_{\mathrm{st}}(E)\,
W_{\mathrm{esc}}(E).
\label{eq:Wesc_weighted}
\end{equation}
This integral is dominated by energies near the saddle point
$E=E_b$, yielding by Laplace's method
\begin{equation}
W_{\mathrm{esc}}
\simeq
C(D)\,
\exp\!\Big(-\frac{U(E_b)-U(E_0)}{D}\Big),
\end{equation}
where $E_0$ is the intrawell minimum and
\[
\Delta U \equiv U(E_b)-U(E_0)
\]
is the activation energy.

\paragraph{Step 6: Identification of the Arrhenius form.}
Defining the slowly varying prefactor
\[
C(D)
=
\frac{\pi\lambda^{2}}{2\hbar}\,
\rho(E_b)\,|X(E_b,\omega)|^{2}
\times (\text{curvature factors}),
\]
we finally obtain
\begin{equation}
\boxed{
W_{\mathrm{esc}}(D)
\simeq
C(D)\,
\exp\!\Big(-\frac{\Delta U}{D}\Big),
}
\label{eq:Warr25_repeat}
\end{equation}
which is Eq.~(25).

In the activated (Kramers) limit the escape is dominated by rare fluctuations that bring the system to the top of the barrier. The net escape rate acquires the Arrhenius-like form (up to a slowly varying prefactor \(C(D)\)):
\begin{equation}
W_{\mathrm{esc}}(D) \simeq C(D)\,\exp\!\Big(-\frac{\Delta U}{D}\Big),
\label{eq:Warr}
\end{equation}
where \(\Delta U\) is the relevant barrier energy (activation energy) and \(C(D)\) is a prefactor which depends weakly on \(D\) (it contains curvatures, damping-dependent factors and power-law dependences on \(D\) that are subdominant in the exponential limit). Equation~\eqref{eq:Warr} is the standard semiclassical reduction of expressions like \eqref{eq:Wesc_integral} once one (i) replaces \(|x(E,E')|^2\) by its semiclassical form (Fourier components of classical motion) and (ii) performs a steepest-descent evaluation of the integral dominated by energies near the barrier top. We therefore adopt \eqref{eq:Warr} as the semiclassical/activated leading-order form.

Now suppose the noise intensity is slowly modulated in time as
\begin{equation}
D(t) = D\,[1 + A\cos(\Omega t)],\qquad |A|\ll 1,
\label{eq:Dmod_repeat}
\end{equation}
the adiabatic/weak-modulation conditions being (i) \(\Omega\) small compared with intrawell relaxation rates so the instantaneous quasi-stationary assumption holds, and (ii) \(A\) small so linearization in \(A\) is justified.

Under these assumptions we replace \(D\mapsto D(t)\) in the activated expression \eqref{eq:Warr} to obtain the instantaneous (adiabatic) escape rate:
\begin{equation}
W_{\mathrm{esc}}(t) \simeq C\big(D(t)\big)\,
\exp\!\Big(-\frac{\Delta U}{D(t)}\Big).
\label{eq:Winst_adiab}
\end{equation}

Because \(A\ll1\) and \(C(D)\) varies slowly with \(D\), expand both prefactor and exponent to first order in \(A\). First, write
\[
\frac{1}{D(t)} = \frac{1}{D[1 + A\cos\Omega t]} = \frac{1}{D}\big(1 - A\cos\Omega t + \mathcal{O}(A^2)\big).
\]
Hence the exponential factor becomes
\begin{align}
\exp\!\Big(-\frac{\Delta U}{D(t)}\Big)
&= \exp\!\Big(-\frac{\Delta U}{D}\big[1 - A\cos\Omega t + \mathcal{O}(A^2)\big]\Big) \nonumber\\
&= \exp\!\Big(-\frac{\Delta U}{D}\Big)\;
\exp\!\Big(\frac{\Delta U}{D}\,A\cos\Omega t + \mathcal{O}(A^2)\Big).
\label{eq:expansion1}
\end{align}
Expanding the second exponential to first order in \(A\) yields
\[
\exp\!\Big(\frac{\Delta U}{D}\,A\cos\Omega t\Big) = 1 + \frac{\Delta U}{D}\,A\cos\Omega t + \mathcal{O}(A^2).
\]
The prefactor \(C(D(t))\) is smooth and contributes only subdominant multiplicative corrections (i.e. \(C(D(t))=C(D)+\mathcal{O}(A)\) but these corrections are small compared with the exponentially large factor when \(\Delta U/D\gg1\)); thus to leading activated order we keep \(C(D)\) fixed.
Combining these linearization and keeping only the leading (exponentially dominant) linear-in-\(A\) contribution we obtain
\begin{equation}
W_{\mathrm{esc}}(t)
\simeq C(D)\,\exp\!\Big(-\frac{\Delta U}{D}\Big)
\Big[ 1 + A\,\frac{\Delta U}{D}\cos\Omega t + \mathcal{O}(A^2,\;C'(D)\text{--terms}) \Big].
\end{equation}
Recognizing \(W_{\mathrm{esc}}^{(0)} = C(D)\exp(-\Delta U/D)\) as the unmodulated escape rate, the result becomes
\begin{equation}
\boxed{%
W_{\mathrm{esc}}(t) \simeq W_{\mathrm{esc}}^{(0)}\Big(1 + A\frac{\Delta U}{D}\cos\Omega t\Big),
}
\label{eq:Wdykman_final}
\end{equation}
which coincides with the linearized modulated form given in Eq.~(6) of Dykman \emph{et al.} (1992).
\section{Results and discussions}
\paragraph{Implications from the derivation in the previous section}
\begin{itemize}
\item In the semiclassical limit, summing over densely spaced states converts this expression into the classical Kramers escape rate, with $D$ (or $\gamma$) corresponding to the noise strength that determines the diffusion over the potential barrier.
\item The leading dependence of the transition rate on the drive amplitude is $W_{m\to n}\propto\lambda^{2}$.
\end{itemize}
\paragraph{Summary of the approximations}
\begin{itemize}
  \item The replacement of the quantum sum by an energy integral and the steepest-descent evaluation yield the activated Arrhenius factor \(e^{-\Delta U/D}\); this requires the semiclassical, high-barrier limit \(\Delta U/D\gg1\).
  \item The adiabatic substitution \(D\mapsto D(t)\) requires the modulation frequency \(\Omega\) to be small compared with intrawell relaxation rates.
  \item Linearization in \(A\) is valid when \(|A|\ll1\) and \(A(\Delta U/D)\) is small enough to neglect higher-order corrections.
\end{itemize}
\paragraph{Justifications on the approximations.}
\begin{enumerate}
  \item The key semiclassical ingredients are: replacing the sum over quantum transitions by an energy integral; expressing matrix elements in semiclassical form (Fourier components of classical motion); and performing a steepest-descent evaluation that yields the Arrhenius exponential \(\exp(-\Delta U/D)\). The prefactor \(C(D)\) can be computed (Kramers prefactor) if required; it does not alter the linear-in-\(A\) modulation factor arising from the exponent.
  \item The adiabatic substitution \(D\mapsto D(t)\) requires \(\Omega\) small compared to intrawell equilibration rates; beyond this regime nonadiabatic corrections alter the simple form. The derivation assumes the adiabatic/instantaneous substitution \(D\mapsto D(t)\) is valid: i.e., the modulation frequency \(\Omega\) is small compared with the inverse relaxation time of intrawell fluctuations so that the intrawell distribution remains quasi-stationary while the noise intensity slowly varies. This is the regime discussed in Dykman \emph{et al.}. 
\item The linear expansion in \(A\) is justified when \(|A|\ll 1\) and \(A(\Delta U/D)\) is small enough that higher-order terms are negligible.
\item The identification \(W_{mn}^{(0)}\propto\exp(-\Delta U_n/D)\) is the standard activated (Kramers) form; more detailed prefactors (Kramers prefactor depending on curvatures and damping) may be retained multiplicatively if desired, but they do not affect the linear-in-\(A\) modulation factor derived above.
	
\end{enumerate}

The central result of this work is the demonstration that the quantum transition rate induced by periodic driving in a nonlinear system reduces, under controlled semiclassical and weak--noise approximations, to an activated form identical in structure to the classical Kramers escape rate. Starting from the microscopic Hamiltonian formulation and first--order time--dependent perturbation theory, the noise--averaged transition rate between quantum states is obtained as
\begin{equation}
W_{mn}^{\mathrm{eff}}
=
\frac{\pi\lambda^{2}}{2\hbar^{2}}
|x_{mn}|^{2}\,
\delta_{\gamma}(\omega_{mn}-\omega),
\qquad
\gamma=\frac{D}{2\hbar^{2}},
\end{equation}
where $\delta_{\gamma}$ is a Lorentzian arising from stochastic phase averaging and $D$ is the noise intensity.

In the semiclassical regime, where the level spacing is small compared to the noise--induced broadening, the sum over final states may be replaced by an integral over energy,
\begin{equation}
W_{\mathrm{esc}}(E)
\simeq
\int dE'\,\rho(E')
\frac{\pi\lambda^{2}}{2\hbar^{2}}
|x(E,E')|^{2}
\delta_{\gamma}\!\left(\frac{E'-E}{\hbar}-\omega\right).
\end{equation}
Using the standard semiclassical correspondence between matrix elements and classical Fourier components,
$|x(E,E')|^{2}\to |X(E,\omega')|^{2}$ with $\omega'=(E'-E)/\hbar$,
and exploiting the smoothness of $\rho(E')$ and $|X(E,\omega')|^{2}$ on the scale set by $\gamma$, the frequency integral can be evaluated explicitly. The resulting energy--resolved transition rate depends parametrically on the classical dynamics at fixed energy.

Weak noise induces slow diffusion in energy space, and the stationary energy distribution satisfies a Fokker--Planck equation whose solution is
\begin{equation}
P_{\mathrm{st}}(E)\propto \exp\!\left(-\frac{U(E)}{D}\right),
\end{equation}
where $U(E)$ is an effective potential determined by classical phase--space transport. The total escape rate is therefore given by
\begin{equation}
W_{\mathrm{esc}}
=
\int dE\,P_{\mathrm{st}}(E)\,W_{\mathrm{esc}}(E),
\end{equation}
which is a Laplace--type integral. In the weak--noise limit $D\to 0$, this integral is evaluated by the method of steepest descent, yielding
\begin{equation}
W_{\mathrm{esc}}(D)
\simeq
C(D)\,
\exp\!\left(-\frac{\Delta U}{D}\right),
\end{equation}
where $\Delta U=U(E_b)-U(E_0)$ is the effective activation barrier and $C(D)$ is a slowly varying prefactor. This establishes a direct correspondence between the quantum transition rate and the classical Kramers rate, with the crucial distinction that $\Delta U$ is not a static potential barrier but an action--like quantity arising from chaotic phase--space transport.

The validity of this reduction rests on three controlled approximations: the semiclassical limit $\hbar\to0$ ensuring a dense spectrum, weak noise such that energy diffusion is slow compared to intrawell dynamics, and adiabatic driving allowing phase averaging to be performed independently of classical transport. Within this regime, chaotic dynamics generated by the interplay of nonlinearity and periodic driving dominates energy transport and provides the intermediate channels responsible for tunneling. Suppression of tunneling under specific driving conditions corresponds mathematically to an increase of the effective barrier $\Delta U$, providing a unified interpretation of chaos--assisted tunneling and coherent destruction of tunneling.

\section{Conclusions and Outlook}

We have shown that tunneling in a driven nonlinear quantum system admits a precise semiclassical description in which quantum transition rates reduce to classical activated escape rates of Kramers type. The derivation proceeds from microscopic transition amplitudes to an effective rate governed by an action barrier determined by classical phase--space transport. Chaos enters the problem through the nonintegrability induced by periodic driving and controls the exponential scaling of the tunneling rate.

This work provides a mathematically consistent framework linking quantum tunneling, chaotic dynamics, and classical activation. It clarifies the role of chaos as an enabling mechanism for tunneling and places chaos--assisted tunneling and coherent destruction of tunneling within a single large--deviation structure. Extensions to higher--dimensional systems, stronger noise, and many--body settings may further illuminate the limits of semiclassical descriptions and the role of chaos in quantum transport.\\
\newpage
{\bf Acknowledgements}:  The author would like to thank Alliance University for providing partial support for carrying out the research work\\

{\bf Declaration of interests:} The sole author has no conflicts of interest to declare. There is no financial interest to report.\\

{\bf Data availability statement:} No data in this publication is to be made available under the study-participant privacy protection clause.\\


\newpage
\begin{thebibliography}{99}

\bibitem{Gutzwiller}
M.~C.~Gutzwiller,
\emph{Chaos in Classical and Quantum Mechanics}
(Springer, New York, 1990).

\bibitem{Ozorio}
A.~M.~Ozorio de~Almeida,
\emph{Hamiltonian Systems: Chaos and Quantization}
(Cambridge University Press, Cambridge, 1988).

\bibitem{Keller}
J.~B.~Keller,
\emph{Geometrical Theory of Diffraction and Related Topics}
(SIAM, Philadelphia, 1980).

\bibitem{Haake}
F.~Haake,
\emph{Quantum Signatures of Chaos}, 3rd ed.
(Springer, Berlin, 2010).

\bibitem{Stoeckmann}
H.-J.~St\"ockmann,
\emph{Quantum Chaos: An Introduction}
(Cambridge University Press, Cambridge, 1999).

\bibitem{DavisHeller}
M.~J.~Davis and E.~J.~Heller,
``Quantum dynamical tunneling in bound states,''
J. Chem. Phys. \textbf{75}, 246--254 (1981).

\bibitem{LinBallentine}
W.~A.~Lin and L.~E.~Ballentine,
``Quantum tunneling and regular and irregular quantum dynamics of a driven double-well oscillator,''
Phys. Rev. A \textbf{45}, 3637--3652 (1992).

\bibitem{TomsovicUllmo}
S.~Tomsovic and D.~Ullmo,
``Chaos-assisted tunneling,''
Phys. Rev. A \textbf{50}, 145--162 (1994).

\bibitem{BohigasTomsovic}
O.~Bohigas, S.~Tomsovic, and D.~Ullmo,
``Manifestations of classical phase space structures in quantum mechanics,''
Phys. Rep. \textbf{223}, 43--133 (1993).

\bibitem{GrossmannDittrich}
F.~Grossmann, T.~Dittrich, P.~Jung, and P.~H\"anggi,
``Coherent destruction of tunneling,''
Phys. Rev. Lett. \textbf{67}, 516--519 (1991).

\bibitem{GrifoniHanggi}
M.~Grifoni and P.~H\"anggi,
``Driven quantum tunneling,''
Phys. Rep. \textbf{304}, 229--354 (1998).

\bibitem{HolthausReview}
M.~Holthaus,
``Floquet engineering with quasienergy bands of periodically driven optical lattices,''
J. Phys. B \textbf{49}, 013001 (2016).

\bibitem{Kramers}
H.~A.~Kramers,
``Brownian motion in a field of force and the diffusion model of chemical reactions,''
Physica \textbf{7}, 284--304 (1940).

\bibitem{FreidlinWentzell}
M.~I.~Freidlin and A.~D.~Wentzell,
\emph{Random Perturbations of Dynamical Systems}, 2nd ed.
(Springer, New York, 1998).

\bibitem{CaldeiraLeggett}
A.~O.~Caldeira and A.~J.~Leggett,
``Path integral approach to quantum Brownian motion,''
Ann. Phys. (N.Y.) \textbf{149}, 374--456 (1983).

\bibitem{DykmanKrivoglaz}
M.~I.~Dykman and M.~A.~Krivoglaz,
``Theory of fluctuational transitions between stable states of a nonlinear oscillator,''
Sov. Phys. JETP \textbf{50}, 30--37 (1979).

\bibitem{DykmanSmelyanskiy}
M.~I.~Dykman, M.~M.~Millonas, and V.~N.~Smelyanskiy,
``Fluctuational transitions between coexisting periodic attractors: The weak-noise limit,''
Phys. Rev. A \textbf{46}, R1713--R1716 (1992).
\bibitem{KuvshinovKuzminPiatrou}
V.~I.~Kuvshinov, A~.V.~ Kuzmin and V.~ A.~Piatrou,
Asymmtric double well system as effective model for the kicked one,
arXiv:1107.0554v1;Chaotic Dynamics(2011)
\bibitem{LinBallentine}
W.~A.~Lin and L.~E.~Ballentine,
``Quantum tunneling and regular and irregular quantum dynamics of a driven double-well oscillator,''
 Phys. Rev. Lett. \textbf{65}, 2927--2920 (1990).

\end{thebibliography}
\end{document}